\documentstyle[12pt]{article}
\renewcommand \baselinestretch{1.4}

\begin{document}

\def\beq{\begin{equation}}
\def\eeq{\end{equation}}
\def\bce{\begin{center}}
\def\ece{\end{center}}
\def\bea{\begin{eqnarray}}
\def\eea{\end{eqnarray}}
\def\ben{\begin{enumerate}}
\def\een{\end{enumerate}}
\def\ul{\underline}
\def\ni{\noindent}
\def\nn{\nonumber}
\def\bs{\bigskip}
\def\ms{\medskip}
\def\wt{\widetilde}
\def\tr{\mbox{Tr}\, }
\def\brr{\begin{array}}
\def\err{\end{array}}




\vspace*{3mm}

\begin{center}

{\LARGE \bf
The conformal anomaly in  N-dimensional spaces having a
hyperbolic spatial section}

\vspace{4mm}

\renewcommand
\baselinestretch{0.8}
\medskip

{\sc A.A. Bytsenko} \\
Department of Theoretical Physics, State Technical University \\
St Petersburg 195251, Russia \\
{\sc E. Elizalde}
\footnote{E-mail: eli@zeta.ecm.ub.es} \\
Center for Advanced Study CEAB, CSIC, Cam\'{\i} de Santa B\`arbara,
17300 Blanes
\\ and Department ECM and IFAE, Faculty of Physics,
University of Barcelona, \\ Diagonal 647, 08028 Barcelona,
Catalonia,
Spain \\
 and \\
{\sc S.D. Odintsov} \footnote{E-mail: odintsov@ecm.ub.es.
On
leave from: Tomsk Pedagogical Institute, 634041 Tomsk, Russia.} \\
Department ECM, Faculty of Physics,
University of  Barcelona, \\  Diagonal 647, 08028 Barcelona, Catalonia,
Spain \\

\vspace{5mm}

{\bf Abstract}

\end{center}

The conformal anomaly for spinors and scalars on a $N$-dimensional
 hyperbolic space is calculated explicitly, by
using zeta-function regularization techniques and the Selberg trace
formula. In the case of conformally invariant spinors and scalars
the results are very much related with those corresponding to a
$N$-dimensional sphere.

\vspace{4mm}

\ni PACS: 04.62.+v, 04.60.-m, 02.30.+g

\newpage

The conformal anomaly \cite{1} plays an important role in different
physical situations. Among others one can mention string theory
\cite{2}, the $C$-theorem and its generalizations \cite{3}, and the
anomaly-induced effective action \cite{4}. Recently, an effective
theory of quantum gravity has been constructed on the basis of
anomaly-induced dynamics \cite{5}. Moreover, the conformal anomaly
(as
well as the total energy-momentum tensor) can be used for the
description of particle production in a gravitational field (see
\cite{6} for a review). The general structure of the conformal
anomaly
in $N$ dimensions has been studied in Ref. \cite{7}. However, the
explicit calculation of the conformal anomaly in a general, curved
spacetime is actually possible only for $N= 2,4,6$ or 8. Already in
the
last case, $N=8$, the calculation is extremely involved.

Under such unfavourable circumstances, in order to obtain  some
information about the conformal anomaly in higher dimension a
possible way is to consider the conformal anomaly in some specified
background. Thus, for instance, the calculation of the conformal
anomaly
on a $N$-dimensional sphere ($N$ even) has been carried out in Ref.
\cite{8} using $\zeta$-function regularization \cite{9} (see
\cite{10}
for a review and detailed list of references). It has been pointed
out
in \cite{8} that the conformal anomaly on a $N$-dimensional sphere
changes sign as $(-1)^{1+N/2}$.

The purpose of this paper is to study the structure of the
conformal anomaly on $N$-dimensional even hyperbolic spaces.
One is led to apply the technique of the Selberg trace formula
in
order to calculate the corresponding zeta function. In this way we
will
obtain a remarkably simple expression for the conformal anomaly on
such a space.

As is well known, for constant conformal transformations the
variation
of the connected vacuum functional $W$ ($W=-\ln Z$, where $Z$ is
the
partition function) can be expressed in terms of the generalized
$\zeta$-function \cite{9} associated with the Laplace-Beltrami
operator
$L_N$ \cite{1}
\beq
\delta W = - \zeta (0| L_N) \ln \mu^2 = \frac{1}{2} \int dV < T_{\mu
\nu}
(x)> \delta g^{\mu \nu} (x),
\label{u1}
\eeq
where $g^{\mu \nu} (x)$ is the metric, $\mu$ is a renormalization
mass
parameter and $<T_{\mu\nu} (x)>$ means that all connected vacuum
graphs
of the stress-energy tensor $T_{\mu\nu} (x)$ are to be included.
Then
Eq. (\ref{u1}) leads to
\beq
<T_\mu^\mu (x) > = \pm V_N^{-1} \zeta (0| L_N^\pm),
\eeq
where for S$^N$: $V_N = 2\pi^{(N+1)/2}\, a^N /\Gamma ((N+1)/2)$,
while
for the compact manifold $H^N/\Gamma$, with $\Gamma$ a co-compact group
of discrete isometries:
 $V_N =V({\cal F}_N )\, a^N$, $a$ being the radius of the
compact space and $V({\cal F}_N)$ the volume of the
fundamental domain ${\cal F}_N$. The sign $+$ (resp. $-$)
corresponds to
integer spin-$s$ fields (resp. spinors).

We will consider the heat-kernel and $\zeta$-function,
respectively,
related with the Laplace-Beltrami operator $L_N = -\nabla^\mu
\nabla_\mu$ acting on fields in $H^N/\Gamma$, with strictly
hyperbolic
group $\Gamma$. The elliptic operator $L_N$ has a pure
discrete spectrum with isolated eigenvalues $\lambda_j$, $j=0,1,2,
\ldots$, of finite multiplicity. With all these assumptions we
shall now apply the Selberg trace formula to the scalar field \cite{11}.
It is convenient to distinguish between the contribution coming from the
identity and the hyperbolic elements of the isometry group
$\Gamma$. Thus, for the case of a scalar field ($s=0$), we have
\bea
\tr \exp (-t L^+_N) &=& V({\cal F}_N) e^{-tb^2} \int_0^\infty dr \,
e^{-tr^2} \Phi_N (r) + \frac{e^{-tb^2}}{(4\pi t)^{1/2}}
\sum_{\{\gamma \}} \sum_{n=1}^\infty  \frac{\chi^n (\gamma )l_\gamma
}{S_N (n; l_\gamma )} \exp \left[ - \frac{(n l_\gamma )^2}{4t}
\right] \nn \\ && \equiv K(t| L^+_N)+ K_h(t| L^+_N),
\label{1}
\eea
and
\bea
\zeta_{H^N/\Gamma} (z|L^+_N) &=& V({\cal F}_N) \int_0^\infty dr \,
(r^2 +b^2)^{-z} \Phi_N (r) + \frac{1}{\sqrt{\pi}\, \Gamma (z)}
\sum_{\{\gamma \}} \sum_{n=1}^\infty  \frac{\chi^n (\gamma )l_\gamma
}{S_N (n; l_\gamma )}  \left(  \frac{2b}{n l_\gamma} \right)^{1/2
-
z}\nn \\ && \times  K_{1/2-z} (nl_\gamma b)  \equiv \zeta (z|
L^+_N)+ \zeta_h(z| L^+_N),
\label{2}
\eea
where $K_\nu (z)$ is the modified Bessel function of the second
kind, $\{ \gamma \} $ denotes the  primitive conjugacy class
determined by the element $\gamma \in \Gamma $, $l_\gamma$ is the
length of the closed geodesic $\gamma$ (which is the same for all
elements in the class $\{ \gamma \} $), $\chi (\gamma ) : \Gamma
\rightarrow S^1$ is a character of $\Gamma$, $S_N (n; l_\gamma)$
some function of $n$ and $l_\gamma$ (see \cite{11} for more
details), and where we have set $a=1$ (in the final results the
dependence on the radius $a$ will be easily restored). Moreover,
\beq
\Phi_N (r) = C_N \mu^+(r,s=0), \ \ \ \ C_N =2^{N-3} \pi^{-(N/2 +1)}
\Gamma (N/2),
\label{3}
\eeq
where $\mu^+ (r,s)$ is the Plancherel measure \cite{12}.

We will now consider the generalization of  these formulas to the
case of symmetric transverse and traceless tensor fields and to
spinors. In addition,
the Selberg trace formula associated with rank-$s$ eigentensors of the
Laplace-Beltrami operator can be written in the same form (\ref{1}),
(\ref{2}), in the appropriate space. For rank-1 eigenvectors, for
example, the character $\chi$ of $\Gamma$ acts  in a  complex linear
space $V$. Therefore, the second term
in Eqs.  (\ref{1}) and (\ref{2}) must be generalized in the sense
that $\chi^n(\gamma ) \rightarrow \tr_V (\chi^n (\gamma ))$, where
$\tr_V$ denotes  the trace operator in the corresponding space $V$
\cite{13}.
Notice, however, that these second terms do not contribute to the
heat-kernel expansion (expansion for $t \rightarrow 0$), nor to the
$\zeta$-function at $z=0$. So we will concentrate our interest on
the functions $K(t|L_N)$ and $\zeta(z|L_N)$ only.

In what follows we will evaluate the trace anomaly for spin-$s$
fields ($s=0,1,2, \ldots$) and Dirac spinors in anti-DeSitter
spacetimes of even dimension $N$. For $N\geq 2$, we have \cite{14}
\bea
&& \mu^\pm (r,s) = \frac{\pi r}{[2^{N-2}\Gamma (N/2)]^2} \tanh [\pi
(r+is)] \sigma^\pm (r,s), \nn \\ &&  \tanh [\pi (r+is)] = \left\{
\brr{ll} \tanh (\pi r) , & s=0,1, \ldots \\ \coth (\pi r), &
s=1/2,3/2, \ldots \err \right.
\label{4}
\eea
where
\beq
\sigma^+ (r,s) = \left[ r^2 + \left( s+ \frac{N-3}{2} \right)^2
\right] \prod_{j=1/2}^{(N-5)/2} (r^2+j^2) \equiv \sum_{k=0}^{N/2 -
1} \beta_{k,N}^+ r^{2k},
\label{5}
\eeq
\beq
\sigma^- (r,1/2) =  \prod_{j=1}^{N/2-1} (r^2+j^2) \equiv
\sum_{k=0}^{N/2 -1} \beta_{k,N}^- r^{2k},
\label{6}
\eeq
the coeficients $\beta^\pm_{k,N}$ being defined by the recasting of
the products into polynomials in $r^2$, in Eqs. (\ref{5}) and (\ref{6}).
For $N=4$
the productory in Eq. (\ref{5}) is omitted and we have: $\beta^+_{1,4}
=1$, $\beta^+_{0,4} = (s+1/2)^2$ and $\beta^-_{0,2} = 1$. The
spectral functions on $H^2$ for spin 0 and 1 are both given
by $\mu^+ (r,0) = \pi r \tanh (\pi r)$.

Substituting (\ref{4})-(\ref{6}) into (\ref{1}) and (\ref{2}), we
obtain
\beq
K(t|L^+_N) = g(s) A(N) e^{-tb^2} \int_0^\infty dr \,
\sum_{k=0}^{N/2-1} \beta^+_{k,N} e^{-tr^2} r^{2k+1} \tanh (\pi r),
\label{7}
\eeq
\beq
K(t|L^-_N) = 2^{N/2} A(N) e^{-tc^2} \int_0^\infty dr \,
\sum_{k=0}^{N/2-1} \beta^-_{k,N} e^{-tr^2} r^{2k+1} \coth (\pi r),
\label{8}
\eeq
\beq
\zeta(z|L^+_N) = g(s) A(N)  \int_0^\infty dr \, \sum_{k=0}^{N/2-1}
\beta^+_{k,N} \frac{ r^{2k+1} \tanh (\pi r)}{(r^2+b^2)^z},
\label{9}
\eeq
and
\beq
\zeta(z|L^-_N) = 2^{N/2} A(N)  \int_0^\infty dr \, \sum_{k=0}^{N/2-
1} \beta^-_{k,N} \frac{ r^{2k+1} \coth (\pi r)}{(r^2+c^2)^z},
\label{10}
\eeq
where
\beq
A(N)= \frac{V({\cal F}_N)}{2^{N-1} \pi^{N/2} \Gamma (N/2)}, \ \ \
\ \ \ g(s)=\frac{(2s+N-3)\ (s+N-4)!}{(N-3)! \, s!}, \ \ \ N \geq 4.
\label{11}
\eeq
Using the identities
\bea
\tanh [\pi (r+is)] = 1 - \frac{2}{1+\exp[2\pi (r+is)]}, &&
\int_0^\infty \frac{dr \, r^{2n-1}}{e^{2\pi r} +1} = \frac{(-1)^{n-
1} (1-2^{1-2n})}{4n} B_{2n},  \nn \\   \int_0^\infty \frac{dr \,
r^{2n-1}}{e^{2\pi r} -1} = \frac{(-1)^{n-1} }{4n} B_{2n}, &&
\label{12}
\eea
where the $B_n$ are Bernoulli's numbers, we obtain
\beq
K(t|L^+_N) =\frac{1}{2} g(s) A(N) e^{-tb^2}  \sum_{k=0}^{N/2-1}
\beta^+_{k,N} \left[ k! t^{-k-1} + (-1)^{k+1} \sum_{l=0}^\infty
\frac{t^l}{l!} \, \frac{1-2^{-2k-2l-1}}{k+l+1} B_{2k+2l+2} \right],
\label{14}
\eeq
\beq
K(t|L^-_N) =2^{N/2-1} A(N) e^{-tc^2}  \sum_{k=0}^{N/2-1}
\beta^-_{k,N} \left[ k! t^{-k-1} + (-1)^{k} \sum_{l=0}^\infty
\frac{t^l}{l! \,(k+l+1)} B_{2k+2l+2} \right].
\label{15}
\eeq

The integrals in Eqs. (\ref{9}) and (\ref{10}) converge for Re $z
>N/2$ and may be analytically continued in the complex $z$-plane by
using the identities (\ref{12}). As a result, we obtain (see also
\cite{10})
 \beq
\zeta (z|L^+_N) =\frac{1}{2} g(s) A(N)  \sum_{k=0}^{N/2-1}
\beta^+_{k,N} \left[ b^{2k-2z+2} B(k+1, z-k-1)-4 \int_0^\infty
\frac{dr \, r^{2k+1}}{(r^2+b^2)^z (e^{2\pi r} +1)} \right],
\label{16}
\eeq
\beq
\zeta (z|L^-_N) =2^{N/2-1} A(N)  \sum_{k=0}^{N/2-1} \beta^-_{k,N}
\left[ c^{2k-2z+2} B(k+1, z-k-1)+4 \int_0^\infty \frac{dr \,
r^{2k+1}}{(r^2+c^2)^z (e^{2\pi r} -1)} \right],
\label{17}
\eeq
in terms of the beta function $B(x,y)=\Gamma (x) \Gamma (y) /\Gamma
(x+y)$. The last terms in Eqs. (\ref{16}) and (\ref{17}) are
analytic in $z$ while the first ones give simple poles for $\zeta
(z| L^\pm_N)$ at $z=1, \ldots, N/2$, in agreement with the general
theory \cite{16}. For further evaluations in terms of Laurent
series, we may rewrite the analytic terms in a more useful form,
taking into account the Mellin transform of the factor $\exp (2\pi
r) \pm 1$:
\beq
[\exp (2\pi r) \pm 1]^{-1} = \frac{1}{2\pi i} \int_{\omega -i
\infty}^{\omega +i \infty} d\sigma \, \zeta^\pm (\sigma ) \Gamma
(\sigma ) (2\pi r)^{-\sigma},
\label{18}
\eeq
where Re $\sigma =\omega$, $\omega > 0$ (resp. $\omega >1$) for
integer (resp. half-integer) $s$, $\zeta^- (\sigma ) = \zeta
(\sigma )$ is the Riemann $\zeta$-function, and $\zeta^+ (\sigma ) =
(1-2^{1-\sigma }) \zeta (\sigma ) =\sum_{n=1}^\infty (-1)^{n-1}
n^{-\sigma} =\eta (\sigma )$, for Re $\sigma >0$, the $\eta$-function.

Using (\ref{18}) in the last terms of (\ref{16}) and (\ref{17}) and
performing the integration over $r$ with the help of the beta
function $B(x,y)$ (notice that, owing to absolute convergence,  the
order of integration over $r$ and $\sigma$ can be interchanged), we
obtain
\bea
\zeta (z|L_N^+ ) &=& \frac{g(s)A(N)}{2\, \Gamma (z)}  \sum_{k=0}^{N/2-1}
\beta^+_{k,N} b^{2k-2z+2} \left[ \Gamma (k+1) \Gamma (z-k-1) \right. \nn
\\
&&- \left. \frac{1}{\pi i}  \int_{\omega -i \infty}^{\omega +i \infty}
d\sigma
\, \zeta^+ (\sigma ) \Gamma (\sigma ) \Gamma (k+1-\sigma /2) \Gamma
(z-k-1+\sigma /2) (2\pi b)^{-\sigma} \right]
\label{19}
\eea
and
\bea
\zeta (z|L_N^- ) &=& \frac{2^{N/2-1}A(N)}{\Gamma (z)}
\sum_{k=0}^{N/2-1} \beta^-_{k,N} c^{2k-2z+2} \left[ \Gamma (k+1)
\Gamma (z-k-1) \right. \nn \\ && + \left. \frac{1}{\pi i}  \int_{\omega
-i \infty}^{\omega
+i \infty} d\sigma \, \zeta^- (\sigma ) \Gamma (\sigma ) \Gamma
(k+1-\sigma /2) \Gamma (z-k-1+\sigma /2) (2\pi c)^{-\sigma}
\right].
\label{20}
\eea
The analytic continuation of the spin-$s$ $\zeta$-function at $z=0$
can be obtained by making  use of  the  asymptotic expansion of
$\Gamma (z)$ at $z=-m$, $m\in$ {\bf N} \cite{17}
\bea
\Gamma (z) &=& \frac{(-1)^m}{m!} \left\{ (z+m)^{-1} + \psi (m+1) +
\frac{z+m}{2} \left[ \frac{\pi^2}{3} + \psi^2 (m+1) - \psi'(m+1)
\right] \right. \nn \\ && + \left. {\cal O} \left( (z+m)^2 \right)
\right\}, \label{21}
\eea
where $\psi (z) = d\ln \Gamma (z)/dz$. Thus, we get
\beq
\zeta (0|L_N^+ ) = \frac{1}{2}g(s)A(N)  \sum_{k=0}^{N/2-1} \frac{(-
1)^{k+1}}{k+1} \beta^+_{k,N} \left[ b^{2k+2} + (1-2^{-2k-1})
B_{2k+2} \right]
\label{22}
\eeq
and
\beq
\zeta (0|L_N^- ) = 2^{n/2-1} A(N)  \sum_{k=0}^{N/2-1} \frac{(-
1)^{k+1}}{k+1} \beta^-_{k,N} \left[ c^{2k+2} - B_{2k+2} \right].
\label{23}
\eeq
Restoring now the dependence on the radius $a$, we obtain the final
result
\beq
<T_\mu^\mu (x)>_+ = \frac{g(s)}{(4\pi)^{N/2}\Gamma (N/2) a^N}
\sum_{k=0}^{N/2-1} \frac{(-1)^{k+1}}{k+1} \beta^+_{k,N} \left[
b^{2k+2} + (1-2^{-2k-1}) B_{2k+2} \right],
\label{24}
\eeq
\beq
<T_\mu^\mu (x)>_- = \frac{1}{(2\pi)^{N/2}\Gamma (N/2) a^N}
\sum_{k=0}^{N/2-1} \frac{(-1)^{k+1}}{k+1} \beta^-_{k,N} \left[
B_{2k+2} - c^{2k+2} \right].
\label{25}
\eeq

Of the two expressions, let us analyze first the one corresponding
to a spinor field. Notice that in the case of a compact manifold,
the $\zeta$-function corresponding to a spinor field is well
defined, since the spectrum of the Dirac operator does not include
the zero point. For massless spinors $c=0$, and we have
\beq
<T_\mu^\mu (x)>_- = \frac{1}{(2\pi)^{N/2}\Gamma (N/2) a^N}
\sum_{k=0}^{N/2-1} \frac{(-1)^{k+1}}{k+1} B_{2k+2} \beta^-_{k,N}.
\label{26}
\eeq
Therefore, in accordance with \cite{8}, for massless spinors we
obtain
\beq
<T_\mu^\mu (x \in H^N )>_- = (-1)^{N/2}
<T_\mu^\mu (x \in S^N )>_-
\label{27}
\eeq
and, for $N=2$, $<T_\mu^\mu (x)>_- = - (12\pi a^2)^{-1}$. As we
see, there appears a very simple relation (only the sign can be
different) between the  conformal anomaly for a spinor on a sphere
and on a hyperbolic space.

For the minimally coupled scalar field of mass $m$, $b^2=\rho^2_N +
a^2m^2$, where $\rho_N = (N-1)/2$. For the conformally invariant
scalar field in $N$ dimensions, we have
\beq
b^2 = \rho_N^2 + \frac{(N-2)a^2}{4(N-1)} R(x),
\label{28}
\eeq
where $R(x) =-N(N-1)/ a^2$ is the scalar curvature. As a
consequence, $b^2 =1/4$ and
\beq
<T_\mu^\mu (x)>_c = \frac{1}{(4\pi)^{N/2}\Gamma (N/2) a^N}
\sum_{k=0}^{N/2-1} \frac{(-1)^{k+1}}{k+1}\beta^+_{k,N} \left[ 2^{-
4k-2} + (1-2^{-2k-1}) B_{2k+2} \right] .
\label{29}
\eeq
In particular,
\beq
<T_\mu^\mu (x)>_c^{N=2} = -\frac{1}{12\pi a^2}, \ \ \ \ \ \
<T_\mu^\mu (x)>_c^{N=4} = -\frac{1}{240\pi^2 a^4},
\label{30}
\eeq
where we have used, for $N=2$, $\beta^+_{0,2} =1, \ g(0)=1$.
The values of the conformal anomaly in the scalar case, for several
dimensions $N=4,6,8,10$ (and $a=1$) are given in Table 1. They are
again the same as the corresponding ones in spherical space, but for a
factor $(-1)^{N/2}$, as in the spinor case.

 \begin{table}

\begin{center}

\begin{tabular}{|c|c|c|c|}
\hline \hline
 $N$ & $<T_\mu^\mu (x)>_c^{hyp}$  &
  $<T_\mu^\mu (x)>_c^{sph}$ & numerical \\ \hline
 \hline 4 & $-\frac{1}{240\pi^2}$
  & $-\frac{1}{240\pi^2}$ & $=-0.00042217$ \\
 \hline 6 & $-\frac{5}{4032\pi^3}$
  & $\frac{5}{4032\pi^3}$ & $=0.00003999$ \\
 \hline 8 & $-\frac{23}{34560\pi^4}$
  & $-\frac{23}{34560\pi^4}$ & $=-6.83211 \cdot 10^{-6}$ \\
 \hline 10 & $-\frac{263}{560880\pi^5}$
  & $\frac{263}{560880\pi^5}$ & $=1.53227 \cdot 10^{-6}$ \\
\hline \hline \end{tabular}
\caption{ {\protect\small Values of the scalar conformal
anomaly on hyperbolic spaces of dimension $N = 4,6,8,10$, respectively
(the radius $a$ has been set equal to 1). The results are
compared with the corresponding ones for the case of a sphere, which are
given in the last two columns. We see that the only difference
is a sign when $N/2$ is odd.  } }

\end{center}

\end{table}

We should draw attention to the fact that, in the case of the
spin-1, vector field theory, for instance,
the Hodge-de Rahm operator $-(d\delta + \delta d) $ acting on co-exact
one-forms is associated with the massless operator $[- \nabla^\mu
\nabla_\mu + (N-1) a^{-2}] g_{\mu\nu}$. The eigenvalues of this
operator are $r^2 + (\rho_N -1)^2$ \cite{14}, and for the Proca
field of mass $m$ we find $b^2= (\rho_N -1)^2 +a^2m^2$.

In conclusion, we have evaluated the conformal anomaly in
spacetimes of arbitrary dimension which possess a compact spatial
section of the form $H^N/\Gamma$. We have restricted ourselves to
the situation where the manifold is smooth and $\Gamma$ is a
discrete subgroup of SO(N,1), acting freely and properly
discontinously on $H^N$. In this case, the `topological terms' on
the right hand side of Eqs. (\ref{1}) and (\ref{2}) do not
contribute to the heat kernel expansion, as we have seen above.
This has simplified the calculations somehow, nevertheless we have also
discussed shortly more general situations. The Selberg trace
formula for the case when the group $\Gamma$ contains elliptic
elements (orbifolds) ---together with the corresponding
$\zeta$-function and the heat-kernel coefficients--- can be found in
\cite{11}. Extension of the above procedure to the explicit
evaluation of the conformal anomaly in such case  seems certainly
feasible.

\vspace{5mm}


\noindent{\large \bf Acknowledgments}

SDO would like to thank the members of the Department ECM, Barcelona
University, for continued hospitality.
This work has been
supported by DGICYT (Spain), project No. PB93-0035,
by CIRIT (Generalitat de Catalunya),
and by RFFR (Russia), project
No. 94-02-03234.

\end{document}